\documentclass[10pt,twocolumn,preprintnumbers,amsmath,amssymb,nofootinbib
,superscriptaddress]{revtex4-2}
\usepackage{amsmath,amssymb,amsfonts}
\usepackage{hyperref}
\usepackage{graphicx}
\usepackage{subfig}
\usepackage{lipsum}
\usepackage{color}

\begin{document}

\title{Post-Newtonian Tests of Gravitational Quantum Field Theory with Spin and Scaling Gauge Symmetry}

\author{Ying-Jian Chen}
\affiliation{International Centre for Theoretical Physics Asia-Pacific (ICTP-AP), University of Chinese Academy of Sciences (UCAS), Beijing 100190, China}
\affiliation{Taiji Laboratory for Gravitational Wave Universe (Beijing/Hangzhou), UCAS, Beijing 100049, China}
    \affiliation{Center for Gravitational Wave Experiment, National Microgravity Laboratory, Institute of Mechanics,
Chinese Academy of Sciences, Beijing 100190, China}

\author{Peng Xu}
\email{xupeng@imech.ac.cn}
\affiliation{Taiji Laboratory for Gravitational Wave Universe (Beijing/Hangzhou), UCAS, Beijing 100049, China}
\affiliation{Center for Gravitational Wave Experiment, National Microgravity Laboratory, Institute of Mechanics,
Chinese Academy of Sciences, Beijing 100190, China}
\affiliation{School of Fundamental Physics and Mathematical Sciences, Hangzhou Institute for Advanced Study, UCAS, Hangzhou 310024, China}
\affiliation{Lanzhou Center of Theoretical Physics, Lanzhou University, Lanzhou 730000, China}

\author{Yue-Liang~Wu}
\email{ylwu@ucas.ac.cn}
\affiliation{International Centre for Theoretical Physics Asia-Pacific (ICTP-AP), University of Chinese Academy of Sciences (UCAS), Beijing 100190, China}
\affiliation{Taiji Laboratory for Gravitational Wave Universe (Beijing/Hangzhou), UCAS, Beijing 100049, China}
\affiliation{Center for Gravitational Wave Experiment, National Microgravity Laboratory, Institute of Mechanics,
Chinese Academy of Sciences, Beijing 100190, China}
\affiliation{School of Fundamental Physics and Mathematical Sciences, Hangzhou Institute for Advanced Study, UCAS, Hangzhou 310024, China}
\affiliation{CAS key laboratory of theoretical Physics, Institute of Theoretical Physics, Chinese Academy of Sciences, Beijing 100190, China}

\begin{abstract}
\noindent
A self-consistent gravitational quantum field theory, with gravitational force treated on the same footing as the other three fundamental interactions, was established recently. 
The gravidynamics predicted by such a theory could lead to important implications, and the comparisons with experimental results may provide us opportunities to test such new approach of gravity based on the framework of the quantum field theory of gauge interactions.  
In this work, we start with the effective field equation of the gravitational quantum field theory, and then solve the perturbative gravigauge field order by order up to the 1st post-Newtonian level under the assumption of a simplified energy-momentum tensor of perfect fluids. 
Having the constraints on the related post-Newtonian parameters from the most up-to-date observational data, the new bound on the combined coupling in the gravitational quantum field theory $|\gamma_G(\alpha_G-\alpha_W/2)|\leq  (2.4\pm30)\times10^{-6} $ is obtained. 
Under such bound, we found that 
the new gravitational quantum field theory successfully passed and found no conflict with the contemporary keynote Solar system experiments of gravity. 
\end{abstract}

\maketitle

\section{Introduction}

Einstein's General theory of Relativity (GR) with his revolutionary perspective of geometrization of fundamental forces and the Quantum Field Theory (QFT) of gauge interactions can be viewed as the two cornerstones of modern physics, while, the union of these two basic frameworks remains an unfinished journey in today's theoretical investigations.
GR, since its beginning in the early 20th century, had passed, systematically, many stringent experimental tests mainly within the weak field limit, especially after the establishments of the so called Dicke-framework of experimental relativity and the Parameterized Post-Newtonian (PPN) formalism since the late 1960s \cite{thorne_theoretical_1971,will_theoretical_1971,will_theoretical_1971-1,ni_theoretical_1972}.
Today, confronted even with the gravitational wave observations of strong field dynamics of curved spacetime from the LIGO-VIRGO-KAGRA collaborations \cite{Krishnendu_2021,perkins_2022,Metha_2023}, GR remains the best fitted theory among all the alternatives of classical theory of gravitation \cite{will_confrontation_2014,will_theory_2018,de_marchi_testing_2020}.
However, challenges in theoretical concerns such as the unification of fundamental forces, information paradoxes and the final fate of black holes, the enigma of spacetime singularities and so on had inevitably led us into the uncharted realm of quantum theory of gravity.
On the other hand, relativistic QFTs in Minkowski spacetime had been verified, with more and more precise experiments, as a successful theoretical framework for the other three fundamental forces. 
The electromagnetic, weak and strong interactions are described by the standard model and are treated as relativistic gauge fields in Minkowski spacetime \cite{tomonaga_relativistically_1946,schwinger_quantum-electrodynamics_1948,feynman_space-time_1949,dyson_s_1949,glashow_partial-symmetries_1961,weinberg_model_1967,gross_ultraviolet_1973,politzer_reliable_1973,fritzsch_advantages_1973,starobinsky_new_1980}.
Such success had strongly motived one to consider the possibility that gravitational force may be treated on the same footing as the other three fundamental interactions and the long-sought quantum theory of gravity could be constructed as a QFT of gauge interaction defined in a flat base spacetime. 

However, such approach is different from the main stream studies of QFTs of gauge theory of gravity in the past half century, see \cite{hehl_general_1976,ivanenko_gauge_1983,hehl_metric-affine_1995} for detailed reviews, which were generally built, following Einstein's ideal of general covariance, based on Riemannian geometry on curved spacetime. 
Considering the critical meaning of the global Poincare group $P(1,3)=SO(1,3)\rtimes T^{1,3}$ in the successful QFT description of the standard model, it has been pointed out that such a global symmetry of the reference or coordinate Minkowski spacetime together with localized internal symmetries, especially the internal translation-like symmetry in relating to $T^{1,3}$,  should be essential in the construction of the QFT of gravity \cite{wu_theory_2015,wu_quantum_2016,wu_foundations_2022,wu_gravidynamics_2023}.   
Such a new approach to the QFT of gravity gauge theory is distinguished from the ideas of general covariance suggested by Einstein, since the general linear group $GL(1,3,R)$ for general coordinates transformations on curved spacetime manifolds does not contain the translation symmetry  $T^{1,3}$ as a subgroup. 
Having this in mind, the main postulates for this new approach of QFT of gravity had been summarized firstly  in refs.\cite{wu_theory_2015,wu_quantum_2016}, (i) a bi-frame spacetime is proposed to describe the QFT of gravity; (ii) The kinematics of all quantum fields obeys the principles of special relativity and quantum mechanics; (iii) The dynamics of all quantum fields is characterized by basic interactions governed by the gauge symmetries; (iv) The action of quantum gravity is to be expressed in the spin-related intrinsic gravigauge spacetime to be coordinate independent and gauge invariant; (v) The theory is invariant not only under the local spin and scaling gauge transformations of quantum fields defined in the gravigauge spacetime, but also under the global Lorentz and scaling as well as translational transformations of coordinates in the flat Minkowski spacetime. 
Based on such postulates, a self-consistent gravitational quantum field theory (GQFT) in a bi-frame spacetime bearing with both the global Poincare symmetry $P(1,3)$ of the flat reference or coordinate spacetime (as the base spacetime) and internal Poincare-type gauge group referred to as inhomogeneous spin gauge symmetry $WS(1,3)=SP(1,3)\rtimes W^{1,3}$  of locally non-coordinate gravigauge spacetime (as the fibers) was established in refs.\cite{wu_quantum_2016,wu_foundations_2022}, where $SP(1,3)\equiv SO(1,3)$ and $W^{1,3}$ correspond to the internal spin gauge symmetry and chiral-type translation-like spin gauge symmetry of fermion fields.  In such a new theory, a bicovariant gravigauge field is identified as the massless graviton within the bi-frame spacetime framework. Notably, the gravitational interactions involving basic fermion fields and spin gauge fields occur primarily via the spin-related gravigauge field rather than the composite gravimetric field.

The detailed gravidynamics and spinodynamics of the GQFT were derived and carefully investigated in ref.\cite{wu_gravidynamics_2023}, where GR can be found as the classical limit of such gravidynamics within the framework of GQFT in the low energy limit. As energy scale $\mu_G$ grows, new effects in gravidynamics predicted by such a theory will gradually deviate from GR, which will become significant important when the energy scale is comparable with the Planck mass scale $\mu_G \sim M_P$, please see \cite{wu_gravidynamics_2023} for details.
This may lead to important implications and shed some new light upon variance exciting topics like the quantum modifications of singularity theorem, quantum inflation in early Universe, the fate of black holes etc., while, on the other hand, the new gravidynamics in this theory will also provide us important clues and opportunities to test such new approach of quantum theory of gravity based on the basic framework of the QFT of gauge interactions. 
Therefore, it is meaningful to systematically compare the predictions of the GQFT with today's experimental results from the precision Solar system tests and the astronomical observations. 

For such a task, the well-known PPN formalism established by Nordtvedt, Will and et al. in the 1970s can serve as a basic but powerful tool \cite{will_confrontation_2014,will_theory_2018}. 
In the PPN formalism, the perturbative metric of a gravitational theory is expanded by orders in terms of linear combinations of the so-called Post-Newtonian (PN) potentials, which are determined by the properties of matter distributions. 
The differences in the matter dynamics in gravitational fields predicted by different theories are encoded in the coefficients (the PN parameters) of these post-Newtonian potentials in the corresponding metrics,  which can be compared with the most up-to-date observation data. 
In this work, the PPN metric solution of the GQFT in the weak field and slow motion limits is derived, and, confronted with today's Solar system experiments, a most up-to-date constraint on the combined key parameter $\gamma_W\equiv\gamma_G(\alpha_G-\alpha_W/2)$ of the GQFT is obtained.
It is found that the new GQFT successfully passed and found no conflict with the contemporary keynote Solar system experiments of gravity.

\section{PN metric solution of GQFT}\label{SecGT}

\subsection{Field equation}
Limited by the length of this work, we will not repeat the detailed derivation of the effective field equation of the GQFT, which can be found in \cite{wu_gravidynamics_2023}.
Here, we summarize the effective field equation that extends Einstein's GR in the large scale and low energy regime of the GQFT and the related definitions used in this work
\begin{equation}
    R_{\mu\nu}-\frac{1}{2}\chi_{\mu\nu}R+\gamma_G\widetilde{R}_{\mu\nu}=-8\pi GT_{\mu\nu}, \label{fieldEq}
\end{equation}
here $R_{\mu\nu}$ denotes the Ricci tensor, $R$ the scalar curvature, $\chi_{\mu\nu}$ serves as the metric, and $T_{\mu\nu}$  the energy-momentum tensor of the matter field. 
The sigh differences $(-,+,+,+)$ and units with $c=1$ are adopted.
The new correction term $\widetilde{R}_{\mu\nu}$ in GQFT  is defined as follows \cite{wu_gravidynamics_2023}: 
\begin{eqnarray}
\widetilde{R}_{\mu\nu}&=&\bar{\nabla}_\rho\bar{F}^\rho_{(\mu\nu)}+\Bigl[\frac{1}{2}\left(\chi_{\mu\lambda}\eta_\nu^\rho+\chi_{\nu\lambda}\eta_\mu^\rho\right)\bar{\chi}_{aa'}^{[\lambda\sigma]\rho'\sigma'}\nonumber\\&&-\frac{1}{4}\hat{\chi}_{\mu\nu}\bar{\chi}_{aa'}^{[\rho\sigma]\rho'\sigma'}\Bigr]F^a_{\rho\sigma}F^{a'}_{\rho'\sigma'},\nonumber\\
\bar{\nabla}_\rho\bar{F}^\rho_{(\mu\nu)}&\equiv&\partial_\rho\bar{F}^\rho_{(\mu\nu)}-\Gamma^\rho_{\rho\lambda}\bar{F}^\lambda_{(\mu\nu)}-\frac{1}{2}\left(\Gamma^\lambda_{\rho\mu}\bar{F}^\rho_{\lambda\nu}+\Gamma^\lambda_{\rho\nu}\bar{F}^\rho_{\lambda\mu}\right),\nonumber\\
\bar{F}^\rho_{\mu\nu}&\equiv&\chi_{\mu\sigma}\chi_\nu^a\bar{\chi}_{aa'}^{[\rho\sigma]\rho'\sigma'}F_{\rho'\sigma'}^{a'},\nonumber\\
F^a_{\rho\sigma}&=&\partial_\rho\chi_\sigma^a-\partial_\sigma\chi_\rho^a,\nonumber\\
\bar{\chi}_{aa'}^{\rho\sigma\rho'\sigma'}&\equiv&\hat{\chi}_c^\rho\hat{\chi}_d^\sigma\hat{\chi}_{c'}^{\rho'}\hat{\chi}_{d'}^{\sigma'}\bar{\eta}_{aa'}^{cdc'd'},\nonumber\\
\bar{\eta}_{aa'}^{cdc'd'}&\equiv&\alpha_G\eta^{cc'}\eta^{dd'}\eta_{aa'}-\frac{1}{2}\alpha_W\left(\eta^{cc'}\eta^d_{a'}\eta^{d'}_a+\eta^{dd'}\eta^c_{a'}\eta^{c'}_a\right).\nonumber
\end{eqnarray}
$\Gamma^\rho_{\mu\nu}$ is the Christoffel symbol, $\eta_{\mu\nu}$ is the Minkowski metric of the global flat base spacetime, $\gamma_G$, $\alpha_G$ and $\alpha_W$ are coupling constants. 
Here $\mu,\nu,\rho,\lambda = 1,2,3,4$ are spacetime indices and $a,b,c,d=1,2,3,4$
the indices for spin-related internal gravigauge spacetime. 
 $\chi_\mu^a$ is the gravigauge bicovariant vector field defined on the global flat Minkowski spacetime and valued in the vector representation of the $SP(1,3)$ symmetry of the local gravigauge spacetime, and $\hat{\chi}^\mu_a$ denotes its inverse. 
The relation between the metric and the bicovariant vector field reads \cite{wu_gravidynamics_2023}
\begin{equation}
    \chi_{\mu\nu}=\chi_\mu^a\chi_\nu^b\eta_{ab} \label{metric}
\end{equation}
It is worth to notice that the GQFT model is related to the so called teleparallel approach to gravity, in which the localized translation symmetry plays the crucial role and the tetrad field arises as the fundamental field variable, please see \cite{Bahamonde,Aldrovandi2012TeleparallelGA} for details. 
In a recent work \cite{WOS:000495071400006}, the PPN expansion of the general formalism of teleparallel gravity was worked out and the corresponding constraints from observations were obtained, and
the comparison between these two models is discussed in the next section.

Without loss of generality, we assume the energy-momentum tensor to take the general form of perfect fluids~\cite{weinberg_gravitation_1972,will_confrontation_2014,will_theory_2018}
\begin{equation}
    T^{\mu\nu}_{f}=(\rho+\rho\Pi+p)u^\mu u^\nu+p\hat{\chi}^{\mu\nu},
\end{equation}
where $\rho$ and $p$ is the mass density and pressure as measured in a local, freely falling, comoving frame, $\Pi$ is the internal energy per unit rest mass, $u^\mu$ is 4-velocity of the fluid element
\begin{equation}
    u^\mu=\left(\frac{1}{\sqrt{1-v^2}},\frac{v^i}{\sqrt{1-v^2}}\right), 
\end{equation}
with $v^i\ \ (i=1,2,3)$ the 3-velocity.
For Solar system experiments within the weak field and slow motion limits, concerning mainly precision measurement techniques like rangings between satellites or between satellites and ground stations, Lunar rangings, clock comparisons in space, satellite gradiometry, VLBI (Very-Long-Baseline radio Interferometry) and so on, the matter sources of stars and planets could generally be modeled as stationary, slowly rotating and almost spherically symmetric compact objects.
Therefore, in this work,
we could simplify the model of the energy-momentum tensor of the matter source, and ignore the internal energy and pressure 
\begin{equation}
    T^{\mu\nu}=\rho u^\mu u^\nu. \label{Tmunu}
\end{equation}

\subsection{PN expansions}

We rewrite the field equation (\ref{fieldEq}) into an equivalent form
\begin{equation}
    R_{\mu\nu}=-8\pi GS_{\mu\nu}-\gamma_G\widetilde{R}_{\mu\nu}+\frac{1}{2}\chi_{\mu\nu}\gamma_G\widetilde{R}, \label{fieldEQ2}
\end{equation}
with
\begin{eqnarray}
    \widetilde{R}&=&\hat{\chi}^{\mu\nu}\widetilde{R}_{\mu\nu},\nonumber\\
        S_{\mu\nu}&\equiv& T_{\mu\nu}-\frac{1}{2}\chi_{\mu\nu}T, \ \ \ \ \  T=\chi_{\mu\nu}T^{\mu\nu}.\nonumber
\end{eqnarray}
This is the basic equation in this work that is to be solved order by order.
To obtain the 1PN metric solution, one need to expand all the related quantities in the above field equation in a self-consistent and balanced way and solve perturbatively the metric components $\chi_{\mu\nu}$. 
According to the Virial theorem for the self-gravitating system, one has
\[
\mathcal{O}(U)\sim \mathcal{O}(v^2),
\]
where $U$ denotes the Newtonian potential,  and from the motions of the matter sources one also has
\[
\partial / \partial t \sim  \vec{v}\cdot \nabla .
\]
Therefore, the bookkeeping of the order of ``smallness'' in the PN expansions can be based on the powers of velocity $v^i$. 
According to the PN formalism, to obtain the balanced orders of the 1PN equation of motions for massive objects 
\[
I=\int \sqrt{\chi_{00}+2\chi_{0i}v^i+\chi_{ij}v^iv^j}dt,
\]
one needs to solve the metric up to the following orders 
\begin{eqnarray}
    \chi_{00}&=&-1+\chi_{00(2)}+\chi_{00(4)}+...,\nonumber\\
    \chi_{0i}&=&\chi_{0i(3)}+...,\nonumber\\
    \chi_{ij}&=&\delta_{ij}+\chi_{ij(2)}+...,\nonumber
\end{eqnarray}
here, similar to Weinberg's notation \cite{weinberg_gravitation_1972}, we use the numbers in parentheses to label the powers of the velocity.

The Ricci tensor and the tensor $S_{\mu\nu}$ should be expanded to the following orders \cite{weinberg_gravitation_1972}
\begin{eqnarray}
   R_{00}&=&R_{00(2)}+R_{00(4)}+...,\nonumber\\
   R_{0i}&=&R_{0i(3)}+...,\nonumber\\
   R_{ij}&=&R_{ij(2)}+...,\nonumber\\
   S_{00}&=&S_{00(0)}+S_{00(2)}+...,\nonumber\\
   S_{0i}&=&S_{0i(1)}+...,\nonumber\\
   S_{ij}&=&S_{ij(0)}+S_{ij(2)}+...,\nonumber
\end{eqnarray} 
therefore, for $\Gamma^i_{00}$, $\Gamma^0_{0i}$ and $\Gamma^i_{jk}$, we have \cite{weinberg_gravitation_1972}
\begin{equation}
\Gamma^\mu_{\nu\lambda}=\Gamma^\mu_{\nu\lambda(2)}+\Gamma^\mu_{\nu\lambda(4)}+...,\nonumber
\end{equation}
and for $\Gamma^i_{0j}$, $\Gamma^0_{00}$ and $\Gamma^0_{ij}$, we have
\begin{equation}
\Gamma^\mu_{\nu\lambda}=\Gamma^\mu_{\nu\lambda(3)}+....\nonumber
\end{equation}
With careful considerations of the correspondences between quantities with upper and lower indices under the PN approximation, we have \cite{weinberg_gravitation_1972}
\begin{eqnarray}
    \chi_{00(2)}&=&-\hat{\chi}^{00(2)}, \  \ \ \ \  \chi_{00(4)}=-\hat{\chi}^{00(4)},\nonumber\\
    \chi_{ij(2)}&=&-\hat{\chi}^{ij(2)}, \  \  \ \ \ \chi_{0i(3)}=\hat{\chi}^{0i(3)},\nonumber    
\end{eqnarray}
and
\begin{eqnarray}
    S_{00(0)}&=&\frac{1}{2}T^{00(0)},\label{s1}\\
    S_{00(2)}&=&\frac{1}{2}(T^{00(2)}-2\chi_{00(2)}T^{00(0)}+\sum_i T^{ii(2)}),\label{s2}\\
    S_{0i(1)}&=&-T^{0i(1)},\label{s3}\\
    S_{ij(0)}&=&\frac{1}{2}\delta_{ij}T^{00(0)}.\label{s4}
\end{eqnarray}

For the gravigauge bicovariant vector $\chi^a_{\mu}$ introduced in the GQFT, the required PN expansions read
\begin{eqnarray}
    \chi_0^0&=&1+\chi_{0(2)}^0+\chi^0_{0(4)}+...,\nonumber\\
    \chi^0_{i}&=&\chi^0_{i(3)}+...,\nonumber\\
    \chi^i_{0}&=&\chi^i_{0(3)}+...,\nonumber\\
    \chi^i_j&=&\delta_{ij}+\chi^i_{j(2)}+\chi^i_{j(4)}+...,\nonumber
\end{eqnarray}
and the PN expansions of $\widetilde{R}$ and the only relevant  00 component of $\widetilde{R}_{\mu\nu}$ are as follows
\begin{eqnarray}
\widetilde{R}&=&\widetilde{R}_{(2)}+\widetilde{R}_{(4)}+...,\nonumber\\
\widetilde{R}_{00}&=&\widetilde{R}_{00(2)}+\widetilde{R}_{00(4)}+....\nonumber
\end{eqnarray}

\subsection{PN metric solution}
Now, for clarity, we denote the metric components as
\begin{eqnarray}
    \chi_{00(2)}&=&h_{00},\ \ \ \ \    \chi_{00(4)}=h_{00(4)},\nonumber\\
        \chi_{0i(3)}&=&h_{0i},\ \ \ \ \ 
        \chi_{ij(2)}=h_{ij},\nonumber
\end{eqnarray}
and according to Eq. (\ref{metric}) and the PN expansion discussed in the previous subsection, the relation between the gravigauge bicovariant vector and the metric components read
\begin{eqnarray}
    -2\chi^0_{0(2)}&=&h_{00},\nonumber\\
        -2\chi^0_{0(4)}&=&h_{00(4)}+\frac{1}{4}h_{00}h_{00},\nonumber\\
       2\chi^i_{i(2)}&=&h_{ii},\nonumber\\
       -2\chi^0_{i(3)}&=&2\chi^i_{0(3)}=h_{0i}.\nonumber
\end{eqnarray}
Given the harmonic gauge condition, and after the tedious calculations and simplifications, we have the PN expansions of the left-hand side of the field equation (\ref{fieldEQ2}).
\begin{eqnarray}
     R_{00(4)}&=&\frac{1}{2}(\partial_i\partial^i h_{00(4)}+\partial_i h_{00}\partial^i h_{00}-\partial_0\partial_0h_{00}-h_{ij}\partial^i\partial^jh_{00}),\nonumber\\
        R_{00(2)}&=&\frac{1}{2}\partial_i\partial^i h_{00},\nonumber\\
        R_{0i(3)}&=&\frac{1}{2}\partial_j\partial^j h_{0i},\nonumber\\
        R_{ij(2)}&=&\frac{1}{2}\partial_k\partial^k h_{ij},\nonumber
        \end{eqnarray}
and the curvature correction terms in the right-hand side of Eq. (\ref{fieldEQ2})
\begin{eqnarray}
    \gamma_G\widetilde{R}_{00(2)}&=&\frac{1}{2}\gamma_W\partial^i\partial_i h_{00},\nonumber\\
        \gamma_G\widetilde{R}_{(2)}&=&\frac{\gamma_W(1+\gamma_W)}{2(1-\gamma_W)}\partial_i\partial^ih_{00},\nonumber\\
        \gamma_G\widetilde{R}_{0i(3)}&=&\frac{\gamma_W}{2}\partial_j\partial^jh_{0i},\nonumber\\
        \gamma_G\widetilde{R}_{ii(2)}&=&\frac{\gamma_W}{3(1-\gamma_W)}\partial_i\partial^ih_{00},\nonumber\\
        \gamma_G\widetilde{R}_{ij(2)}&=&-\frac{\gamma_W}{2(1-\gamma_W)}\partial_i\partial_jh_{00},\nonumber\\
\gamma_G\widetilde{R}_{00(4)}&=&\frac{\gamma_W}{2}\partial_i\partial^i h_{00(4)}+\frac{\gamma_W}{8}\partial_i\partial^ih_{00}^2\nonumber\\
         &&+\frac{\gamma_W(6\Gamma-5)}{4}h_{00}\partial_i\partial^i h_{00}\nonumber\\&&+\frac{\gamma_W(10\Gamma^2-2\Gamma-5)}{8}\partial_i h_{00}\partial^i h_{00},\nonumber\\
        -\gamma_G\widetilde{R}_{(4)}&=&\frac{\gamma_W}{2}\partial_i\partial^ih_{00(4)}+\frac{\gamma_W}{8}\partial_i\partial^ih_{00}^2\nonumber\\&&+\gamma_W\left(2\Gamma^2+\frac{\Gamma}{2}-\frac{3}{4}\right)h_{00}\partial_i\partial^ih_{00}\nonumber\\&&+\gamma_W\left(\frac{15\Gamma^2}{4}-\frac{3\Gamma}{4}-\frac{1}{2}\right)\partial_ih_{00}\partial^ih_{00},\nonumber
    \end{eqnarray}
    with
    \begin{eqnarray}
        \Gamma\equiv\frac{3+\gamma_W}{3-3\gamma_W}, \ \ \ \ 
        \gamma_W\equiv\gamma_G(\alpha_G-\alpha_W/2).\nonumber
\end{eqnarray}
Together with Eq. (\ref{s1}) - (\ref{s4}) and the definitions of the energy-momentum tensor in Eq. (\ref{Tmunu}) 
\begin{eqnarray}
    T^{00(0)}&=&\rho^*, \ \ \ \ \ \
        T^{00(2)}=\frac{1}{2}\rho^* v^2+(2-3\Gamma)\rho^*U,\nonumber\\
        T^{0i(1)}&=&\rho^* v^i,\ \ \ \ \ 
        T^{ij(2)}=\rho^* v^iv^j,\nonumber
\end{eqnarray}
where
\begin{eqnarray}
    \rho^*&=&\rho u^0\sqrt{-\chi},\nonumber\\
   U&=&G^*\int\frac{\rho'^*}{\mid \vec{x}-\vec{x}'\mid}d^3x',\nonumber\\
   G^*&\equiv&\frac{2(1-\gamma_W)}{(2-\gamma_W)(1+\gamma_W)}G .\nonumber
\end{eqnarray}
$\chi$ is the determinant of metric. We substitute the above PN expansions into Eq. (6), and obtain the three equations for the linear solutions of the metric components $h_{00},\ h_{0i}$ and $h_{ij}$
\begin{eqnarray}
   -16\pi G\rho^*&=& (1+\gamma_W)\partial_i\partial^ih_{00}+\frac{1}{2}\partial_i\partial^ih,\label{PNeq00}\\
  16\pi G\rho^* v_i& =&(1+\gamma_W)\partial_j\partial^jh_{0i}, \label{PNeq0i}\\
   0&=&(1+\gamma_W)\partial_k\partial^kh_{ij}-\frac{\gamma_W}{2}\partial_i\partial_jh\nonumber\\&&-\frac{\eta_{ij}}{2}\partial_k\partial^kh\label{PNeqij}
   \end{eqnarray}
   with
\[h\equiv \sum_ih_{ii}-h_{00},\]
and the equation for the nonlinear metric term $h_{00(4)}$
\begin{eqnarray}
    &\frac{2+\gamma_W}{4}\partial_i\partial^ih_{00(4)}-\frac{1}{2}h_{ij}\partial^i\partial^jh_{00}+\frac{\gamma_W}{16}\partial_i\partial^ih_{00}^2\nonumber\\&+\left(\frac{1}{2}-\frac{3}{8}\gamma_W+\frac{1}{8} \gamma_W\Gamma-\frac{5}{8}\gamma_W\Gamma^2\right)\partial_ih_{00}\partial^ih_{00}\nonumber\\&-\left(\gamma_W\Gamma^2-\frac{5\gamma_W}{4}\Gamma+\frac{7}{8}\gamma_W+\frac{\gamma_W}{4}\frac{1+\gamma_W}{1-\gamma_W}\right)h_{00}\partial_i\partial^ih_{00}\nonumber\\&=-4\pi G\rho^*\left(\frac{3}{2}v^2-(2+3\Gamma)U\right).\label{PNnonlinear}
\end{eqnarray}

The equations for the linear terms Eq. (\ref{PNeq00}) - (\ref{PNeqij}) are solved in the first place, and then the nonlinear term can be solved by substituting the linear metric solutions  $h_{00},\ h_{0i}$ and $h_{ij}$ into above Eq. (\ref{PNnonlinear}). 
We summarize the final 1PN metric solutions for GQFT in the form of a PPN metric \cite{will_theory_2018} under harmonic gauge,
\begin{eqnarray}
    \chi_{00}&=& -1+2U-(2\beta+\gamma-1) U^2+2\psi\nonumber\\
    &&+(\gamma+\sigma_1)\ddot{X}+\mathcal{O}(v^6),\label{chi00}\\
    \chi_{0i}&=&  -2\left(\gamma+1\right)V_i +\mathcal{O}(v^5), \label{chi0i}\\
    \chi_{ij}&=& \delta_{ij}(1+2\gamma U)-(1-\gamma)X_{,ij}+\mathcal{O}(v^4),\label{chiij}
\end{eqnarray}
with the related PN potentials \cite{will_theory_2018}.
\begin{eqnarray}
      \Phi_1&=&G^*\int\frac{\rho'^* v'^2}{\mid \vec{x}-\vec{x}'\mid}d^3x',\nonumber\\
    \Phi_2&=&G^*\int\frac{\rho'^*U'}{\mid \vec{x}-\vec{x}'\mid}d^3x',\nonumber\\
     \psi&=&\frac{1}{2}(2\gamma+1+\sigma_2)\Phi_1-(2\beta-\frac{3}{2}+\frac{1}{2}\gamma+\frac{1}{2}\sigma_2)\Phi_2,\nonumber\\
    \ddot{X}&=&-\frac{\partial^2}{\partial t^2}\int\frac{1}{2\pi}\frac{U'}{\mid \vec{x}-\vec{x}'\mid}d^3x',\nonumber\\
    V_i&=&G^*\int\frac{\rho'^* v'_i}{\mid \vec{x}-\vec{x}'\mid}d^3x',\nonumber\\
    X_{,ij}&=&U\delta_{ij}-U_{ij},\nonumber\\
    U_{ij}&=&G^*\int\frac{\rho'^* (x-x')_i(x-x')_j}{\mid \vec{x}-\vec{x}'\mid^3}d^3x',\nonumber
    \end{eqnarray}

    \begin{table*}
    \centering
    \begin{tabular}{|c|c|c|c|}
    \hline
           GQFT PN  parameters&Effects& Constraints of PN parameters& Constraints of $\gamma_W$\\
         \hline
           $\gamma-1$&deflection of light & $(-0.8\pm1.2)\times10^{-4}$ & $(-0.8\pm1.2)\times10^{-4}$\\
         \hline
           $\gamma-1$&time delay & $(2.1\pm2.3)\times10^{-5}$ & $(2.1\pm2.3)\times10^{-5}$\\
         \hline
           $\gamma-1$&Mercury precession & $(-0.3\pm2.5)\times10^{-5}$ & $(-0.3\pm2.5)\times10^{-5}$\\
         \hline 
           $\beta-1$&Mercury precession & $(0.2\pm2.5)\times10^{-5}$ & $(2.4\pm30)\times10^{-6}$\\\hline
    \end{tabular}
    \caption{The constraints on the PN parameters for GQFT and the coupling parameter $\gamma_W$ of GQFT.}
    \label{tab:my_label}
\end{table*}
The related non-zero PN parameters of the GQFT read,
    \begin{eqnarray}
    \gamma&=&\frac{1}{1-\gamma_W},\label{gamma}\\
        \beta&=&\frac{36-24\gamma_W+33\gamma_W^2-41\gamma_W^3}{18(1-\gamma_W)^2(2+\gamma_W)},\label{beta}
        \end{eqnarray}
with 2 additional correction parameters
\begin{eqnarray*}
    \sigma_1&=&-\frac{\gamma_W(37\gamma_W^2-66\gamma_W+111)}{9(2+\gamma_W)(\gamma_W-1)^2},               \\
     \sigma_2&=&-\frac{\gamma_W(55\gamma_W^2-75\gamma_W+102)}{9(2+\gamma_W)(\gamma_W-1)^2} ,              
\end{eqnarray*}
We see that only one parameter $\gamma_W = \gamma_G(\alpha_G-\alpha_W/2)$ from the effective GQFT appears in the PN metric solution, which is the combination of the three coupling constants of the theory. 
As expected, with $\gamma_W \rightarrow 0$, the PPN metric of the GQFT in Eq. (\ref{chi00}) - (\ref{chiij})  will return to that of GR \cite{will_theory_2018}.

\section{Experimental constraints on GQFT}\label{SecConstraint}

In the standard PPN formalism, the PN parameters may have certain physical meanings or are related to specific physical phenomenons in curved spacetime.
In GQFT the conservation law of energy-momentum is preserved \cite{wu_quantum_2016,wu_gravidynamics_2023}.
Therefore, in this work, the related PN parameters $\alpha_3$ and $\zeta_m\ (m=1,2,3,4)$ reflecting the possible violations of the conservation of energy-momentum and the internal pressure and energy in matter source modelings are ignored for the sake of simplicity.  
The PN parameters $\xi$ and $\alpha_j\ (j=1,2,3)$ and the related PN potentials reflecting the preferred frame effects are also ignored here, since the GQFT satisfies the basic postulate that the theory is independent of the choices of coordinate systems or reference frames \cite{wu_quantum_2016}. 
The formal full PPN expansion of the GQFT is out of the scope of this letter and will be left for future works.

The important non-zero Eddington-Robertson-Schiff parameters $\gamma$ and $\beta$ are derived for the GQFT in this work, please see Eq. (\ref{gamma}) and (\ref{beta}), where $\gamma$ measures the amount of curvature of space produced by mass and $\beta$ measure the amount of nonlinearity $\sim U^2$ in superposition of gravity.
There are 3 additional correction parameters $\sigma_1,\sigma_2$ in this theory that modify slightly the (relative) magnitudes of the corresponding metric terms, that can not be restored into the standard form of the PPN metric. 


The strongest constraints on these parameters base on the up-to-date experimental results are summarized in Table \ref{tab:my_label}.
In the weak field and slow motion limits, null geodesics are more sensitive to curvature of space.
Through the classical test of deflections of light rays carried out firstly by Eddington a century ago \cite{dyson_ix_1920}, the value of $\gamma$ could be directly determined \cite{will_theory_2018}. 
Today, with the help of modern technologies and methods in precision measurements, especially the VLBI method \cite{shapiro_measurement_2004},  the resolution in measuring light deflection angles had been greatly improved. 
In recent years, VLBI observations of quasars and radio galaxies had established a precision reference frame for astrometry, which had improved the measurements of light deflection angles to milliarcsecond level. 
The careful analyses of the VLBI data through 2010 had yielded the constraint $\gamma-1=(-0.8\pm1.2)\times10^{-4}$ \cite{lambert_determining_2009,lambert_improved_2011}. 
Measurements of the Shapiro time delay of light, another classical test in experimental relativity,  can also put constraints on $\gamma$. 
A strong result $\gamma-1=(2.1\pm2.3)\times10^{-5}$ was reported in \cite{bertotti_test_2003}  from the Doppler tracking data of the Cassini spacecraft while it was on its way to Saturn.
On the other hand, observations of motions of massive objects could set constraints on other PN parameters. 
As the third classical test in experimental relativity, the explanation of the anomalous perihelion shift of Mercury's orbit \cite{einstein_explanation_1915} provided an estimate for the combination $(2+2\gamma-\beta)/3$ \cite{will_theory_2018}. 
Combining the results from all the other available data, the even stronger bounds $\gamma-1=(-0.3\pm2.5)\times10^{-5}$ and $\beta-1=(0.2\pm2.5)\times10^{-5}$ were obtained \cite{fienga_inpop10a_2011,verma_use_2014,fienga_numerical_2015}.
At last but not least, the test of Lense-Thirring effect or frame-dragging effect through observations of satellite laser rangings \cite{ciufolini_dragging_2007,ciufolini_test_2016,ciufolini_lares_2023} and GP-B mission \cite{overduin_spacetime_2015,everitt_gravity_2015} could set the rather weaker constraints on $\gamma$.    


From the above discussions, the experimental constraints on the key Eddington-Robertson-Schiff parameters $\gamma$  and $\beta$ will further set bounds on the combined coupling parameter $\gamma_W$ of GQFT in the weak field and slow motion limit.
The strongest bound is from the studies of anomalous perihelion shift of Mercury's orbit,  which gives $|\gamma_W|=|\gamma_G(\alpha_G-\alpha_W/2)|\leq (2.4\pm30)\times10^{-6}$, which improves a former constraint $|\gamma_W\leq \mathcal{O}(10^{-5})|$ \cite{Gao2024}. 
Such a bound implies that the mass scale of spin gauge field should be around or below the energy scale of grand unification theory. 
Moreover, one can further conclude that, under such bound, the new GQFT had successfully passed today's most stringent and keynote tests in experimental relativity, including precision measurements of light deflection, Shapiro time delay, perihelion shift of Mercury's orbit, Lense-Thirring effect of satellite orbit, frame-dragging precession of orbiting gyroscope etc. 
For the PPN approximation of a general teleparallel gravity theory, it is found that \cite{WOS:000495071400006} the theory does not exhibit any violations in Einstein's equivalence principle and conservation law of energy-momentum. 
It contains only two nontrivial PN parameters $\gamma$ and $\beta$, and, in some sense similar to our results in the last section, the deviations in these parameters from GR depend only on a unique parameter, see again \cite{WOS:000495071400006} for detailed discussions.
This manifests, in some sense, the potential links between the GQFT model considered in this work and general teleparallel gravity.

\section{Concluding remarks}\label{SecConlusion}

In this work, we start with the effective field equation of the GQFT, which can be viewed as an generalization of Einstein's gravidynamics in the large scale and low energy regime of GQFT. 
Then, the perturbative metric of the GQFT is carefully solved order by order up to the 1st PN level, under the assumption of a simplified energy-momentum tensor of perfect fluids. 
Having this result, one can directly compare with the experimental constraints on the related PN parameters, and then set the new and stringent bound on the combined coupling in the GQFT $|\gamma_G(\alpha_G-\alpha_W/2)|\leq  (2.4\pm30)\times10^{-6}$ in the large scale and low energy regime. 
Satisfying such bound, the GQFT successfully passes and finds no conflict with today's keynote Solar system experiments of gravitational theories, including precision measurements of light deflection, Shapiro time delay, perihelion shift of Mercury's orbit, Lense-Thirring effect of satellite orbit, frame-dragging precession of orbiting gyroscope etc. 
Moreover, with future improvements in the precisions of these experiments, the PPN approximation and expansion of the GQFT developed in this work may provide us potential clues and opportunities to tests such new approach of quantum theory of gravity based on the basic framework of the QFT of gauge interactions.

\section*{Acknowledgements}
This work is supported by the National Key Research and Development Program of China, Grant No. 2021YFC2201901 and No.2020YFC220150, International Partnership Program of the Chinese Academy of Sciences, Grant No. 025GJHZ2023106GC, and the National Science Foundation of China (NSFC) under Grants No.~12347103 and No.~11821505.

\appendix




\bibliography{PN_test_GQFT}






\end{document}